\documentclass[%
 aps,
 pra,
 reprint,
 superscriptaddress,
 floatfix,
 amsmath,amssymb,
 nofootinbib,
]{revtex4-2}

\usepackage{graphicx}
\usepackage{dcolumn}
\usepackage{bm}
\usepackage{hyperref}
\usepackage{braket}
\usepackage{subcaption}
\usepackage{booktabs}
\usepackage{xcolor}

\begin{document}

\title{{Tunable M-Ary Quantum Secure Direct Communication via Correlation-Histogram Modulation}}

\author{Todd M.\ W.\ Hodges}
\email{Todd.Hodges@aexp.com}
\affiliation{American Express Co., Phoenix, Arizona, USA}

\date{\today}

\begin{abstract}
We propose a two-way entanglement-based quantum secure direct communication (QSDC) protocol in which information is encoded as the value of a continuously tunable parameter and decoded from empirically estimated joint-outcome histograms over windows of detected pairs. The protocol uses a fixed two-qubit polarization Hilbert space and only standard coincidence-based polarization measurement at the receiver, with the symbol alphabet enlarged not by increasing the carrier Hilbert-space dimension but by exploiting temporal integration over a window of pairs to resolve distinguishable joint-outcome probability distributions. Under an idealized model in which the source is maximally entangled, the encoding operation is strictly local, and the channel imprints no parameter-dependent signature on the traveling photon, the reduced state of the traveling photon is maximally mixed and independent of the encoded parameter. Consequently, passive interception of the in-flight subsystem alone yields no information about the encoded message. We further develop a quantitative trace-distance framework for bounding any parameter-dependent leakage that may arise from non-ideal channel and device effects in practical implementations. The protocol introduces protocol-level design parameters not available in fixed-alphabet two-way QSDC, specifically a runtime-tunable alphabet size and native support for continuous-time analog modulation, which may be advantageous in operating environments where channel conditions vary on protocol-relevant timescales.
\end{abstract}

\maketitle

\section{Introduction}\label{sec:intro}

Quantum cryptography and communications is most widely associated with quantum key distribution (QKD), where quantum channels help to establish the basis for a shared secret key later used by classical cryptographic methods. In QKD, Alice and Bob exchange quantum signals and then use an authenticated classical channel for sifting, parameter estimation, error correction, privacy amplification, and ultimately communication. Security stems from the idea that any eavesdropping interaction that yields information about unknown quantum states also induces disturbances that are statistically detectable, enabling bounds on an adversary's knowledge from observed error statistics~\cite{BennettBrassard1984,ShorPreskill2000,Scarani2009}. Prepare-and-measure protocols such as BB84 infer secrecy from the quantum bit error rate in randomly chosen bases~\cite{BennettBrassard1984,ShorPreskill2000}, while entanglement-based protocols such as E91 and BBM92 frame security in terms of distributed entanglement and integrity of bipartite correlations or Bell-inequality violations~\cite{Ekert1991,BennettBrassardMermin1992,Acin2007,Scarani2009}.

A distinctly different family of quantum communication protocols aims not to distribute a key, but to transmit encoded information directly over a quantum channel. This family of protocols is typically referred to as quantum secure direct communication (QSDC) or deterministic quantum communication and is conceptually differentiated from QKD by placing the encoded information in quantum carriers rather than in a post-processed classical key~\cite{BostromFelbinger2002,DengLongLiu2003,Pan2024EvolutionQSDC}. While QKD can tolerate partial exposure of raw quantum data so long as privacy amplification removes residual leakage, QSDC protocols generally seek to ensure that the message itself is inaccessible to an adversary under the assumed access model and test structure. Many QSDC schemes employ two-way quantum communication in which a quantum system is transmitted from Alice to Bob and returned (or vice versa), enabling Bob to modulate correlations that can be accessed only by joint measurements at the receiver. The ping-pong protocol is a widely studied entanglement-based deterministic example~\cite{BostromFelbinger2002}, and subsequent analysis highlighted subtleties in lossy channels and the importance of explicit threat models and robust testing~\cite{Wojcik2003PingPongEavesdrop,cai2003ping}. Related two-way families, including the Lucamarini--Mancini (LM05) protocol, motivated further theoretical and experimental study~\cite{LucamariniMancini2005,Lucamarini2006LM05Security,Beaudry2013TwoWaySecurity,Kumar2008TwoWayImplementation}, while block-based QSDC protocols such as Deng--Long--Liu illustrate randomized subset testing before decoding on the remaining systems~\cite{DengLongLiu2003}.

This work is positioned within the two-way entanglement-based direct-communication paradigm. We propose an entanglement-correlation modulation protocol in which Alice prepares polarization-entangled photon pairs, retains one photon locally, and sends the partner to Bob. Bob applies a controllable local unitary $U(\theta)$ on the received photon and returns it to Alice, who performs joint measurements and decodes symbols from reconstructed correlation histograms over the states $\{\ket{HH},\ket{HV},\ket{VH},\ket{VV}\}$. Unlike two-way schemes that encode symbols via a finite set of operations mapping to a discrete Bell-state alphabet, the present approach supports distribution-shaped encoding. Specifically, symbols correspond to distinguishable joint-outcome probability vectors estimated over finite temporal windows, enabling discretized M-ary alphabets and, in principle, continuous-time modulation $\theta(t)$. In an idealized model, the reduced state of the traveling photon is maximally mixed and independent of $\theta$. Consequently, measurement of the traveling subsystem alone yields no information about the encoded information. The encoded message resides in bipartite correlations accessible only by joint measurements performed by Alice. Since real channels and devices are not ideal, we also present a quantitative framework for bounding any $\theta$-dependent leakage that may arise from non-ideal effects.

{The contribution of this work is methodological. We develop a two-way QSDC protocol in which information is encoded as the value of a continuously tunable parameter and decoded from empirically estimated joint-outcome histograms over windows of detected pairs. The protocol operates entirely within a fixed two-qubit polarization Hilbert space, typically realized in practice via a source non-degenerate in frequency or path to allow deterministic separation of the two photons, and uses only standard coincidence-based polarization measurement at the receiver. This stands in contrast to prior approaches that enlarge the symbol alphabet by enlarging the Hilbert-space dimension of the carrier through numerous methods~\cite{Wang2005HighDim,Cozzolino2019HighDim,HighDimSciRep2024}. The information capacity of the resulting alphabet is determined by how many joint-outcome probability distributions can be reliably distinguished from each other via finite-sample estimates, which depends on the number of detected pairs per symbol window and therefore on both integration time and the pair-detection rate. The protocol does not, however, require any expansion of the carrier Hilbert space or any new measurement modalities beyond standard coincidence detection. As far as the author is aware, this places the protocol in an operating regime that has not been explored in the prior two-way QSDC literature.}

The protocol-level flexibility introduced by this approach is most useful in operating environments where channel conditions vary on protocol-relevant timescales. Free-space and satellite-based quantum links represent examples of such environments. In satellite-to-ground quantum communication, atmospheric turbulence causes channel transmittance to fluctuate on millisecond-to-second timescales, and a range of channel-adaptive techniques have been developed to mitigate these fluctuations, including adaptive real-time selection of high-transmittance intervals~\cite{Vallone2015ARTS}, prefixed-threshold real-time selection~\cite{Wang2018PrefixedThreshold}, and orbit subdivision in continuous-variable satellite QKD~\cite{Dequal2021SatCVQKD}. These techniques adapt receiver-side processing or segment-level protocol parameters to instantaneous channel conditions, but they operate within a fixed symbol alphabet. The present protocol adds a complementary axis of adaptation by allowing the alphabet size itself to be reduced during low-transmittance intervals (with the integration window extended to preserve symbol distinguishability) and expanded during high-transmittance intervals to recover bits per symbol. This adaptability is one example of the broader class of operational benefits that follow from the protocol's design flexibility.

The remainder of this paper is organized as follows. Section~\ref{sec:protocol} presents the protocol and develops the analytical structure of the encoded states. Section~\ref{sec:ideal_eve} establishes the ideal-model result that the traveling subsystem alone carries no information about the encoded parameter. Section~\ref{sec:window_scaling} develops the finite-window decoding analysis, the resolvable alphabet size as a function of the number of detected pairs per symbol window, and the corresponding per-pair and per-second information rates in comparison to fixed-alphabet two-way QSDC. Section~\ref{sec:leakage} then provides a quantitative framework for bounding $\theta$-dependent leakage that may arise from non-ideal channel and device effects, and Section~\ref{sec:discussion} discusses the relationship of this work to prior continuous-parameter, high-dimensional, and discrete-alphabet QSDC schemes and concludes with directions for future work.

\section{Protocol: Encoding and Histogram-Based Decoding}\label{sec:protocol}

In this section, we use the linear polarization basis states $\ket{H},\ket{V}$ that correspond to the computational basis states $\ket{0}$ and $\ket{1}$. In this polarization basis, the maximally entangled Bell states are
\begin{align}
\ket{\Phi^{\pm}} &= \frac{\ket{HH}\pm\ket{VV}}{\sqrt{2}},&
\ket{\Psi^{\pm}} &= \frac{\ket{HV}\pm\ket{VH}}{\sqrt{2}}.
\end{align}

We propose a two-way entanglement-based protocol that encodes information into the empirical observation probabilities of joint measurements on a bipartite system. For this derivation, we use a simple communication system model with two nodes (Alice and Bob) and a bi-directional link. Alice retains one photon of each pair, labeled $q_0$. The partner photon, sent from Alice to Bob, is labeled $q_1$; the photon returned from Bob to Alice is labeled $q_1'$.

At a high-level, Alice generates polarization-entangled photon pairs, and then sends one photon from each pair to Bob ($q_1$). Bob modulates a parameterized unitary $U(\theta)$ on the traveling subsystem ($q_1$) and returns it to Alice ($q_1'$), who decodes a message symbol from the resulting two-photon histogram estimated over a block of $N$ returned pairs (corresponding to an integration-time window). This creates an opportunity for discretized multi-level alphabets and continuous-time modulation via $\theta(t)$. In practical implementations, the polarization-entangled pair would typically be generated via a source non-degenerate in frequency or path to allow deterministic separation of the two photons. This auxiliary degree of freedom acts only as a sorting label and does not carry encoded information, so the information capacity of the protocol continues to reside in the two-qubit polarization Hilbert space.

Alice may choose which Bell state to generate when her polarization-entangled photon pairs are generated. For this example, Alice will choose $\ket{\Psi^+}$:
\begin{equation}\label{eq:Pol_Bell_State}
    \ket{\psi}=\ket{\Psi^+}=\frac{1}{\sqrt{2}}(\ket{HV}+\ket{VH}).
\end{equation}

As $q_1$ traverses Bob's node, he applies a half-wave plate (HWP) rotation to $q_1$. An ideal HWP implements (up to global phase)
\begin{equation}
U_{\mathrm{HWP}}(\theta)=
\begin{pmatrix}
\cos(2\theta) & \sin(2\theta)\\
\sin(2\theta) & -\cos(2\theta)
\end{pmatrix},
\label{eq:hwp_matrix}
\end{equation}
which corresponds to the state transformations
\begin{align}
\ket{H} &\mapsto \cos(2\theta)\ket{H}+\sin(2\theta)\ket{V}, \label{eq:hwp_H}\\
\ket{V} &\mapsto \sin(2\theta)\ket{H}-\cos(2\theta)\ket{V}. \label{eq:hwp_V}
\end{align}
The post-operation bipartite state is now $\theta$-dependent and may be expressed as
\begin{equation}
\ket{\psi(\theta)}=(I\otimes U_{\mathrm{HWP}}(\theta))\ket{\Psi^+}.
\end{equation}
Let $c\equiv \cos(2\theta)$ and $s\equiv \sin(2\theta)$. Then
\begin{align}
\ket{\psi(\theta)}
&=\frac{1}{\sqrt{2}}\left[s\left(\ket{HH}+\ket{VV}\right)+c\left(\ket{VH}-\ket{HV}\right)\right].
\label{eq:psi_hv_form}
\end{align}
Using $\ket{\Phi^+}=(\ket{HH}+\ket{VV})/\sqrt{2}$ and $\ket{\Psi^-}=(\ket{HV}-\ket{VH})/\sqrt{2}$, we obtain
\begin{equation}
\ket{\psi(\theta)}=\sin(2\theta)\ket{\Phi^+}-\cos(2\theta)\ket{\Psi^-}.
\label{eq:psi_bell_superposition}
\end{equation}

\paragraph{Histogram-based decoding and coincidence processing.}
After Bob returns $q_1'$, Alice performs joint measurement on $(q_0,q_1')$ in the $\{\ket{H},\ket{V}\}$ basis and estimates the joint-outcome probabilities from time-tagged detection events. In a typical implementation, a polarization analysis stage maps $\ket{H}$ and $\ket{V}$ to separate single-photon detection channels from which two-photon outcome counts are inferred by forming coincidence histograms from the recorded arrival-time tags (equivalently, estimating second-order correlation statistics as a function of relative delay). This coincidence/correlation approach is standard in quantum optics and photon-counting measurements. It is formalized by Glauber's photodetection theory and the associated correlation functions, and is widely treated in canonical texts. In practice, it is implemented using time-correlated single-photon counting (TCSPC) or time-tagging electronics to build $R(\tau)$-type estimators from discrete detection events~\cite{Glauber1963PRL,MandelWolf1995,Becker2005TCSPC,Spiess2023ClockSync}. For completeness, we write idealized point-process correlation estimators below, which connect directly to the histogram entries. Let the time-tagged detection events be modeled as
\begin{equation}\label{eq:point_process_sigs}
    x_H(t)=\sum_i\delta(t-t_i),\qquad
    y_V(t)=\sum_j\delta(t-s_j),
\end{equation}
where $t_i$ and $s_j$ are the photon arrival times at each detector. We then define the cross-correlation estimator as
\begin{equation}\label{eq:theo_cc}
    R_{HV}(\tau)=\sum_{i,j}\delta(\tau-(s_j-t_i)).
\end{equation}
In realistic systems, relative path delay and detector jitter broaden coincidences into approximately Gaussian peaks centered near $\tau=\pm \tau_0$; integrating over these peaks yields the total $\ket{HV}$ and $\ket{VH}$ counts, respectively. Auto-correlation estimators,
\begin{equation}\label{eq:ac_HH}
    R_{HH}(\tau)=\sum_{i,j}\delta(\tau-(t_j-t_i)),
\end{equation}
\begin{equation}\label{eq:ac_VV}
    R_{VV}(\tau)=\sum_{i,j}\delta(\tau-(s_j-s_i)),
\end{equation}
provide the $\ket{HH}$ and $\ket{VV}$ counts from the corresponding delayed peaks. These counts define the empirical histograms used for decoding.

\begin{figure}[t]
\centering
\includegraphics[width=0.95\columnwidth]{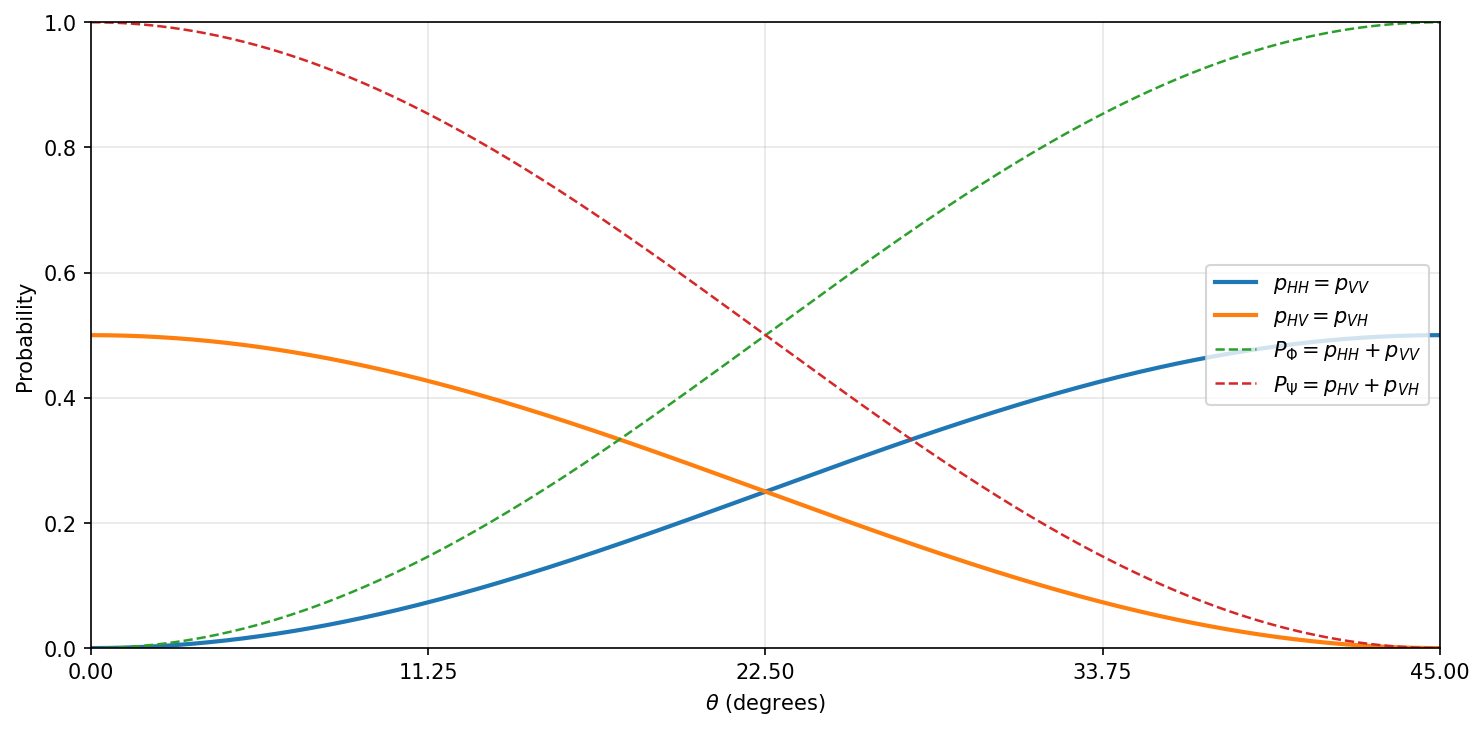}
\caption{Probabilities of two-photon polarization-entangled observables and corresponding Bell state observables as a function of HWP angle $\theta$.}
\label{fig:state_prob_relation}
\end{figure}

\paragraph{Encoding rule and alphabet size.}
From Eq.~\eqref{eq:psi_hv_form}, the $\theta$ dependence induces paired changes in joint-outcome probabilities. Specifically, $\ket{HH}$ and $\ket{VV}$ vary together, and $\ket{HV}$ and $\ket{VH}$ vary together. More formally, the probability of observing the $\ket{\Phi^+}$ and $\ket{\Psi^-}$ states are
\begin{equation}
    P_{\Phi}=p_{HH}+p_{VV}=\sin^2(2\theta),
\end{equation}
\begin{equation}
    P_{\Psi}=p_{HV}+p_{VH}=\cos^2(2\theta)
\end{equation}
respectively, where $\{p_{HH}, p_{HV}, p_{VH}, p_{VV}\}$ represent the corresponding two-photon state probabilities. From this relationship, a simple binary rule of the form
\begin{equation}\label{eq:binary_encoding}
    B=
    \begin{cases}
    0 & P_{\Phi}>P_{\Psi}\\
    1 & P_{\Phi}<P_{\Psi}
    \end{cases},
\end{equation}
could be used for encoding purposes. However, the continuous nature of the $\theta$ variable creates a more robust symbol environment. Figure~\ref{fig:state_prob_relation} shows the relationship between $P_{\Phi}$ and $P_{\Psi}$ in terms of the relationship between $\{p_{HH}, p_{HV}, p_{VH}, p_{VV}\}$.

\begin{figure}[t]
\centering
\includegraphics[width=0.85\columnwidth]{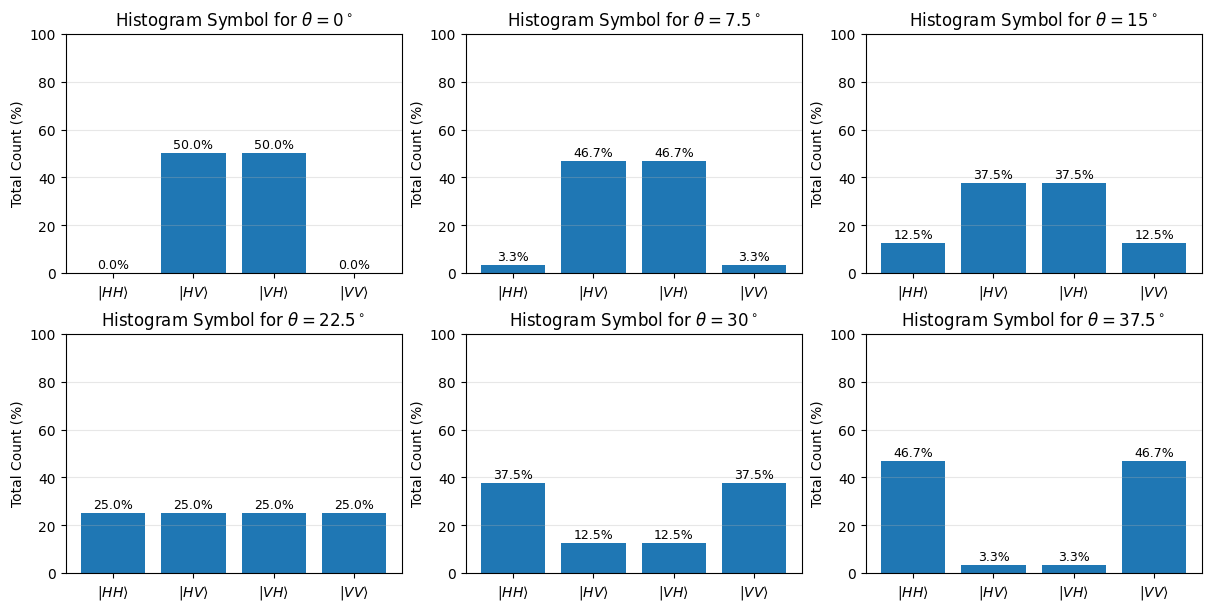}
\caption{Example of a six-symbol alphabet as represented by observable histograms constructed over a time window. Each histogram, which represents a unique set of observation probabilities, corresponds to a specific value of $\theta$.}
\label{fig:histo_lib}
\end{figure}

From the continuous set of pair-wise probability relationships that are a function of $\theta$ as shown in Fig.~\ref{fig:state_prob_relation}, Bob may form a discretized set $\{\theta_m\}_{m=1}^M$ to form an $M$-ary alphabet (Fig.~\ref{fig:histo_lib}), or vary $\theta$ continuously as a function of time $\theta(t)$ to encode an analog waveform into the evolving correlation distribution. In practice, symbol resolution is limited by finite-sample uncertainty in the estimated histogram, noise, and loss.

\section{Eavesdropper Observations in the Ideal Model}\label{sec:ideal_eve}

Assume an eavesdropper (Eve) can access only the traveling subsystem in the quantum channels ($q_1$ or $q_1'$), while Alice's retained subsystem $q_0$ is never transmitted. In the idealized model, interception of $q_1'$ provides no information about the encoded parameter.

To illustrate this, we note that the two-qubit density operator in the ordered basis $\{\ket{HH},\ket{HV},\ket{VH},\ket{VV}\}$ is
\begin{equation}
    \rho(\theta) = \ket{\psi(\theta)}\bra{\psi(\theta)} = \frac{1}{2}
    \begin{pmatrix}
    s^2 & -sc & sc & s^2\\
    -sc & c^2 & -c^2 & -sc\\
    sc & -c^2 & c^2 & sc\\
    s^2 & -sc & sc & s^2
    \end{pmatrix},
\end{equation}
with $s\equiv \sin(2\theta)$ and $c\equiv \cos(2\theta)$. As a result, the reduced state of the traveling photon as calculated via partial trace is
\begin{equation}\label{eq:reduced_density}
    \rho_{q_1'}(\theta) = \mathrm{Tr}_{q_0}\!\left[\rho(\theta)\right]=\frac{I}{2},
\end{equation}
which corresponds to the information that is accessible to Eve. Therefore, all single-photon measurement statistics of $q_1'$ are independent of $\theta$ in the ideal model as the encoded information resides entirely in the bipartite correlations accessible only to joint measurement of $(q_0,q_1')$.

\section{Finite-Window Decoding, Alphabet Size, and Scaling}\label{sec:window_scaling}

In the proposed protocol, the need to construct symbol histograms from finite time windows of $N$ detected pairs creates a tradeoff between symbol distinguishability and operating cost. Specifically, a larger window leads to a larger $N$, which improves symbol distinguishability but increases the integration time before a symbol decision can be reported and the amount of classical memory and computation required to accumulate and process the time-tagged events. The pair-detection rate, the optics, and the measurement architecture are unchanged. In this section we provide simple, standard estimates for symbol resolution versus window size.

\subsection{Histogram estimation noise and resolvable symbol spacing}
Within one window, Alice estimates the joint outcome probabilities
$\hat{\mathbf{p}} = (\hat{p}_{HH},\hat{p}_{HV},\hat{p}_{VH},\hat{p}_{VV})$
from counts $\mathbf{n}=(n_{HH},n_{HV},n_{VH},n_{VV})$ with $\sum_i n_i=N$.
Under a multinomial model,
\begin{align}
\mathbb{E}[\hat{p}_i] &= p_i, \label{eq:multinomial_mean}\\
\mathrm{Var}[\hat{p}_i] &= \frac{p_i(1-p_i)}{N}, \label{eq:multinomial_var}\\
\mathrm{Cov}[\hat{p}_i,\hat{p}_j] &= -\frac{p_i p_j}{N},\quad i\neq j. \label{eq:multinomial_cov}
\end{align}

For the present encoding, the paired probabilities $P_{\Phi} = p_{HH} + p_{VV} = \sin^2(2\theta)$ and $P_{\Psi} = p_{HV} + p_{VH} = \cos^2(2\theta) = 1 - P_{\Phi}$ depend on a single scalar parameter, so the full multinomial distribution over the four joint outcomes is determined by $P_\Phi$ alone. Aggregating counts into the two complementary subsets yields a binomial distribution for the merged count, $n_{\Phi}=n_{HH}+n_{VV}\sim\mathrm{Binomial}(N,P_{\Phi})$, and the corresponding estimator and its variance are

\begin{equation}
\hat{P}_{\Phi}=\frac{n_{\Phi}}{N},\qquad
\mathrm{Var}[\hat{P}_{\Phi}]=\frac{P_{\Phi}(1-P_{\Phi})}{N}.
\label{eq:binomial_Pphi}
\end{equation}
As a design choice, we adopt a conservative separability criterion that requires the probabilities of two neighboring symbols ($\theta_m,\theta_{m+1}$) to differ by at least $k$ standard deviations of the estimated $\hat{P}_{\Phi}$:
\begin{equation}
\Delta P_{\Phi}\equiv \left|P_{\Phi}(\theta_{m+1})-P_{\Phi}(\theta_m)\right|
\;\gtrsim\; k\,\sqrt{\frac{P_{\Phi}(1-P_{\Phi})}{N}},
\label{eq:deltaP_spacing}
\end{equation}
where $\Delta P_\Phi$ is the probability spacing and $k$ is a tunable margin parameter that sets the desired error-rate buffer under a Gaussian (large-$N$) approximation. Since $P_{\Phi}(\theta)=\sin^2(2\theta)$, a first-order Taylor expansion about a representative operating point $\bar{\theta}$ (e.g., $\bar{\theta}\approx(\theta_m+\theta_{m+1})/2$) gives, for sufficiently small $\Delta\theta\equiv|\theta_{m+1}-\theta_m|$,
\begin{equation}
\Delta\theta \approx \frac{\Delta P_{\Phi}}{\left|\frac{d}{d\theta}\sin^2(2\theta)\right|_{\theta=\bar{\theta}}}
= \frac{\Delta P_{\Phi}}{2|\sin(4\bar{\theta})|}.
\label{eq:theta_spacing_local}
\end{equation}
This shows that symbol spacing in $\theta$ is generally nonuniform. Specifically, resolution is poorest near $\sin(4\bar{\theta})\approx 0$ (e.g., $\bar{\theta}\approx 0^\circ,\,45^\circ$ in the range $[0^\circ,45^\circ]$) and best near $|\sin(4\bar{\theta})|\approx 1$ (e.g., $\bar{\theta}\approx 22.5^\circ$ in $[0^\circ,45^\circ]$, and more generally $\bar{\theta}\approx 22.5^\circ+45^\circ l$ for any integer $l$).

\subsection{Implications for alphabet design}
Equations~\eqref{eq:binomial_Pphi}--\eqref{eq:theta_spacing_local} together define a design space for the protocol operator. The free parameters are the window size $N$, which is set by latency and integration-time budgets, the separability margin $k$, which is set by the acceptable symbol error rate, and the placement of the discretized angles $\{\theta_m\}$ within $[0,\pi/2]$. Given any two of these, the third is determined by the requirement that all neighboring symbol pairs satisfy Eq.~\eqref{eq:deltaP_spacing}. In practice, additional non-idealities (loss, background counts, detector jitter, drift) modify the effective $N$ and the observed variances. However, the $1/\sqrt{N}$ scaling remains the dominant factor. Security-side multi-copy scaling considerations for an interceptor are addressed separately in Sec.~\ref{sec:leakage}.

\subsection{Information rate and comparison to discrete-alphabet two-way QSDC}
The information conveyed per symbol in the present scheme is $\log_2 M$ bits, where $M$ is the size of the chosen alphabet, and each symbol is decoded from a window of $N$ detected pairs. The corresponding per-pair information rate is therefore
\begin{equation}
R_{\mathrm{hist}} = \frac{\log_2 M}{N} \quad \text{bits per detected pair},
\label{eq:R_hist}
\end{equation}
where $M$ and $N$ are independent design parameters constrained by Eq.~\eqref{eq:deltaP_spacing}. By contrast, in fixed four-state Bell-state-encoded two-way QSDC, each pair carries the symbol directly via a Bell-state assignment, giving up to $R_{\mathrm{BSM}} = 2$ bits per pair under ideal Bell-state measurement.

The proposed scheme is therefore not intended to compete with fixed-alphabet encoding on a bits-per-pair basis. Instead, the scheme operates in a different region of the design space by using many photon pairs to construct one symbol drawn from a flexible alphabet, while fixed-alphabet encoding uses one pair per symbol drawn from a fixed alphabet. The fundamental tradeoff is therefore between per-pair rate and alphabet flexibility, and the operator chooses the operating point based on the application.

The corresponding bits-per-second rate is
\begin{equation}
R_{\mathrm{time}} = \frac{\log_2 M}{T_{\mathrm{win}}} = \frac{r_{\mathrm{pair}} \log_2 M}{N} \quad \text{bits per second},
\label{eq:R_time}
\end{equation}
where $T_{\mathrm{win}} = N/r_{\mathrm{pair}}$ is the integration-time window required to acquire $N$ detected pairs at an effective pair rate of $r_{\mathrm{pair}}$. At a pair rate of $r_{\mathrm{pair}} = 10^7$ pairs per second, a window of $N = 10^3$ pairs supporting $M = 21$ symbols (for $k = 3$) corresponds to $T_{\mathrm{win}} = 100\,\mu\text{s}$ and $R_{\mathrm{time}} \approx 4.4\times 10^4$ bits per second. The same source running fixed four-state Bell-state encoding at the ideal $2$ bits per pair would deliver $2\times 10^7$ bits per second, so on a bits-per-second basis the fixed-alphabet scheme retains a nearly three-orders-of-magnitude advantage in this example. Both rate figures are idealized upper bounds that assume lossless channels, perfect detection, and no security overhead. Realized rates in any practical implementation are reduced by channel loss, detector inefficiency, finite-size statistics, and any test rounds devoted to security verification. The motivation for adopting the present scheme rather than a higher-rate fixed-alphabet alternative is therefore not its rate, but the operational benefits discussed in Sec.~\ref{sec:discussion}, including runtime alphabet tunability, simplified receiver architecture, and native support for analog modulation. For supporting hardware capability references on SPDC pair rates and modern single-photon detection, see~\cite{Jabir2017BrightSPDC,park2025ultrabright,Zhang2019SNSPDArray,Korzh2020SNSPDJitter}.

\section{Information Leakage Under Non-Ideal Channels}\label{sec:leakage}

The ideal-model conclusion of $\rho_{q_1'}(\theta)=I/2$ relies on a maximally entangled source, a strictly local unitary $U(\theta)$ applied by Bob, and a channel/device that does not imprint $\theta$-dependent signatures onto the traveling subsystem alone. In realistic systems, imperfections such as polarization-dependent loss, optical hardware variance, and $\theta$-dependent back-reflections can produce a reduced state of
\begin{equation}
\rho_{q_1'}^{\mathrm{(real)}}(\theta) \neq \frac{I}{2}.
\end{equation}
Consequently, it is possible for these imperfections to create $\theta$-dependent mapping between symbols and traveling subsystem states for which $\rho_{q_1'}^{\mathrm{(real)}}(\theta_m)\neq\rho_{q_1'}^{\mathrm{(real)}}(\theta_n)$. In such cases, an adversary with access only to the traveling photon may be able to infer partial information about the encoding. The remainder of this section provides a quantitative, experimentally checkable framework to bound such leakage.

\subsection{Which noise processes create leakage}
Not all noise breaks the ideal-model obfuscation property. Unital noise maps satisfy $I/2\mapsto I/2$ and therefore do not create $\theta$-dependent leakage on the traveling subsystem. Leakage arises when the effective channel/device map depends on $\theta$. Examples include polarization-dependent loss that varies with the HWP angle, basis-dependent detection efficiencies, or device back-reflections that carry classical side information correlated with $\theta$. Additionally, because communication is conditioned on detection events, $\theta$-dependent loss can induce $\theta$-dependent conditional states and hence leakage even when the pre-loss marginal is ideally maximally mixed.

As a minimal parametric model, suppose the traveling photon experiences a $\theta$-dependent polarization filter characterized by effective transmissivities (or detection efficiencies) $\eta_H(\theta)$ and $\eta_V(\theta)$ for the $\ket{H}$ and $\ket{V}$ components, respectively:
\begin{equation}
\rho_{q_1'}^{\mathrm{(real)}}(\theta)\propto F_\theta \left(\frac{I}{2}\right) F_\theta^\dagger,
\label{eq:pd_loss}
\end{equation}
where
\begin{equation}
F_\theta = \begin{pmatrix}\sqrt{\eta_H(\theta)} & 0 \\ 0 & \sqrt{\eta_V(\theta)}\end{pmatrix},
\label{eq:filter_op}
\end{equation}
followed by renormalization. If $\eta_H(\theta)\neq \eta_V(\theta)$ and the imbalance varies with $\theta$, then $\rho_{q_1'}^{\mathrm{(real)}}(\theta)$ becomes biased in the $\{\ket{H},\ket{V}\}$ basis and can leak information about the encoded symbol.

\subsection{Pairwise distinguishability and the Helstrom bound}
For a discretized $M$-ary alphabet $\{\theta_m\}_{m=1}^M$ with symbol probabilities $p_m$, Eve faces an $M$-ary quantum state discrimination problem on the ensemble $\{p_m,\rho_m\}$ where $\rho_m \equiv \rho_{q_1'}^{\mathrm{(real)}}(\theta_m)$. In the absence of measurement information, a natural blind baseline is to guess according to the symbol probabilities, yielding $P_{\mathrm{succ}}^{\mathrm{blind}}=\sum_m p_m^2$, which reduces to $1/M$ under uniform probabilities.

While the optimal $M$-ary success probability depends on the full ensemble and measurement, a conservative leakage diagnostic is to bound binary distinguishability induced by any symbol-dependent change in the traveling-subsystem state. In particular, for any two candidate symbols $(m,n)$, the optimal single-copy success probability for deciding ``$\rho_m$ versus $\rho_n$'' is given by the Helstrom limit,
\begin{equation}
P_{\mathrm{succ}}^{\ast}(m,n) = \frac{1}{2}\left(1 + D(\rho_m,\rho_n)\right),
\label{eq:helstrom}
\end{equation}
where the trace distance is defined as
\begin{equation}
D(\rho_m,\rho_n) \equiv \frac{1}{2}\left\|\rho_m-\rho_n\right\|_1.
\label{eq:trace_distance}
\end{equation}
This pairwise quantity should be interpreted as a best-case binary distinguishability metric (useful as a stress test), not as an assumption that Eve knows the transmitted symbol is restricted to two possibilities. In the ideal model $\rho_m=\rho_n=I/2$ for all $(m,n)$, so $D=0$ and $P_{\mathrm{succ}}^{\ast}(m,n)=1/2$ (random guessing within that binary sub-task). Thus $D(\rho_m,\rho_n)$ provides a direct baseline of leakage per intercepted photon at the pairwise level.

A worst-case pairwise summary metric is
\begin{equation}
D_{\max} \equiv \max_{m\neq n} D(\rho_m,\rho_n),
\label{eq:dmax}
\end{equation}
which upper-bounds the distinguishability of the most separated symbol pair on the traveling subsystem and provides a simple criterion for ruling out strong single-arm leakage in calibration.

We use pairwise trace distance as the leakage diagnostic because of its operational clarity and its direct connection to the Helstrom bound. Tighter information-theoretic upper bounds on Eve's accessible information for the full $M$-ary problem are given by the Holevo quantity $\chi(\{p_m,\rho_m\}) = S(\sum_m p_m \rho_m) - \sum_m p_m S(\rho_m)$~\cite{Holevo1973}, where $S(\cdot)$ denotes the von Neumann entropy. The Holevo quantity reduces to a small value whenever $D_{\max}$ is small, so the pairwise metric used here is conservative relative to the tight bound and is sufficient for calibration-based ruling-out of strong leakage.

\subsection{Worked example and many-copy scaling}
To illustrate the scaling behavior of the leakage diagnostic, we consider a worked example using the parametric model of Eq.~\eqref{eq:pd_loss}. Specifically, we set $\eta_H(\theta) = 1 + \epsilon\cos(4\theta)$ and $\eta_V(\theta) = 1 - \epsilon\cos(4\theta)$, which produces a $\theta$-dependent imbalance whose periodicity matches that of the legitimate signal $P_{\Phi}(\theta) = \sin^2(2\theta)$. Under this model, the reduced state $\rho_{q_1'}^{\mathrm{(real)}}(\theta)$ is diagonal in the $\{\ket{H},\ket{V}\}$ basis with diagonal entry $p_H(\theta) = \eta_H(\theta)/(\eta_H(\theta) + \eta_V(\theta))$, so the pairwise trace distance reduces to $D(\rho_m,\rho_n) = |p_H(\theta_m) - p_H(\theta_n)|$. For this diagonal family the optimal Helstrom measurement is the classical $H$ versus $V$ count-rate test, and Eve's success probability after $N_E$ intercepted photons may be approximated, for sufficiently large $N_E$, as $P_{\mathrm{succ}}(N_E) \approx \Phi(\Delta p_H \sqrt{N_E})$, where $\Phi(\cdot)$ is the standard normal cumulative distribution function and $\Delta p_H$ is the worst-case probability separation over the symbol set. The Gaussian-CDF form is a large-$N_E$ approximation to the exact $N_E$-copy trace distance, which is available in closed form for diagonal qubit states~\cite{NielsenChuang2010}, and is sufficient for the illustrative purpose here.

More generally, because symbols are inferred from histograms over finite windows, any realistic interception strategy will involve multiple photons associated with the same encoded symbol. If Eve obtains $N_E$ independent intercepted copies corresponding to a fixed symbol, then pairwise discrimination is governed by $\rho_m^{\otimes N_E}$ and $\rho_n^{\otimes N_E}$, and the relevant distinguishability becomes $D(\rho_m^{\otimes N_E},\rho_n^{\otimes N_E})$, which generally increases with $N_E$. Even small single-copy leakage ($D\ll 1$) can accumulate into stronger distinguishability when many photons are available within a symbol window. The single-copy metrics in Eqs.~\eqref{eq:helstrom}--\eqref{eq:dmax} should therefore be interpreted as per-photon leakage diagnostics, while symbol-level leakage should be evaluated at the intended window size and expected interception fraction. We emphasize that the worked example above is chosen for analytical clarity rather than as a general bound. For other non-ideal channel models, the reduced states need not be diagonal in the measurement basis, and the optimal Helstrom measurement may not coincide with the classical count-rate test. The qualitative scaling, namely that single-copy leakage accumulates with $N_E$ even when small, is expected to be general.

\begin{figure*}[t]
\centering
\includegraphics[width=0.95\textwidth]{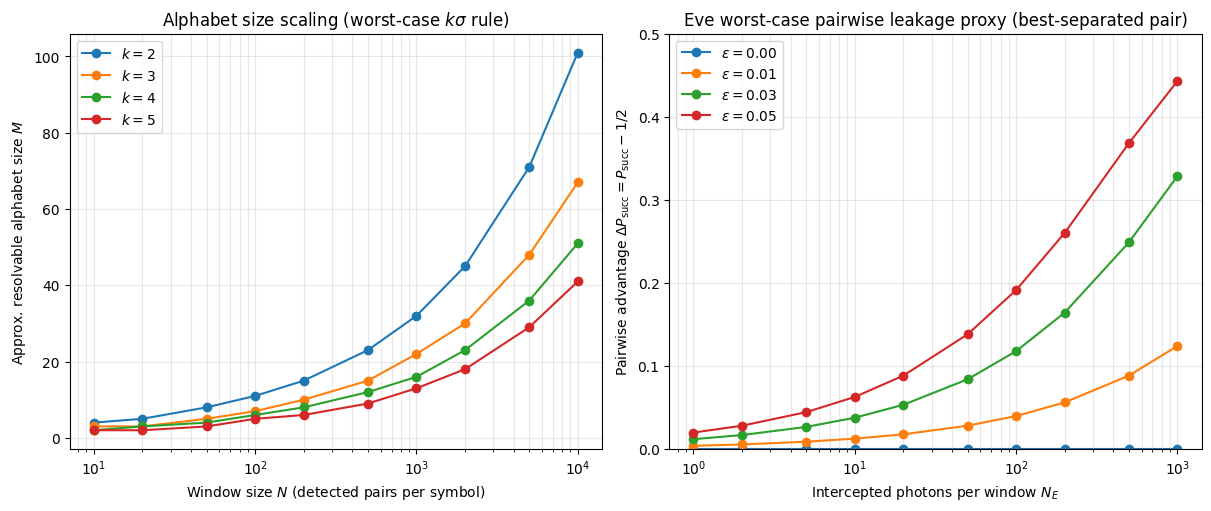}
\caption{Finite-window decoding and worst-case pairwise leakage scaling. \textbf{Left:} A conservative estimate of resolvable alphabet size $M$ versus window size $N$ using a worst-case $k\sigma$ separability rule (Sec.~\ref{sec:window_scaling}). \textbf{Right:} A worst-case pairwise single-arm leakage proxy shown as Eve's advantage $\Delta P_{\mathrm{succ}} = P_{\mathrm{succ}}-1/2$ for distinguishing the most-separated symbol pair, under a simple $\theta$-dependent polarization bias model with strength $\epsilon$ (Sec.~\ref{sec:leakage}). These curves do not represent $M$-ary symbol decoding; they provide a conservative diagnostic of how any residual $\theta$-dependent signature could accumulate with the number of intercepted photons per window $N_E$.}
\label{fig:alphabet_vs_N_and_eve_adv}
\end{figure*}

Figure~\ref{fig:alphabet_vs_N_and_eve_adv} summarizes these finite-window scaling considerations, showing the resolvable alphabet size as a function of $N$ and the worst-case pairwise single-arm advantage as a function of $N_E$ under the simple bias model.

\subsection{From distinguishability to information}
The pairwise distinguishability $D(\rho_m,\rho_n)$ can be translated into an information-theoretic leakage bound by modeling Eve's optimal binary discrimination as a binary symmetric channel with crossover probability $p_e(m,n) = (1 - D(\rho_m,\rho_n))/2$. The corresponding mutual information per intercepted photon is bounded above by
\begin{equation}
I_{\mathrm{Eve}}(m,n) \;\lesssim\; 1 - h_2\!\left(\tfrac{1}{2}(1 - D(\rho_m,\rho_n))\right) \text{ bits},
\label{eq:mutinfo_bound}
\end{equation}
where $h_2(\cdot)$ is the binary entropy. For multi-level alphabets, the same construction applied pairwise across the symbol set bounds per-photon symbol confusability in bits.

\subsection{Practical bounding and calibration}
A practical implementation can bound leakage by estimating $\rho_{q_1'}^{\mathrm{(real)}}(\theta_m)$ for the employed symbol set $\{\theta_m\}$ using calibration runs and standard single-qubit tomography or monitoring taps. The worst-case pairwise distinguishability $D_{\max}$ in Eq.~\eqref{eq:dmax} upper-bounds Eve's single-copy pairwise discrimination advantage across the alphabet and provides a quantitative criterion for selecting discretization levels of $\theta$ that keep traveling-subsystem leakage below a target threshold. When symbol decisions are formed from $N$ detected pairs per window, the same workflow can be applied at the window level by analyzing the resulting $N$-copy (or empirical) statistics, thereby directly bounding leakage at the operating point.

\section{Discussion and Conclusions}\label{sec:discussion}

\subsection{Positioning and intended use cases}
The present protocol is not designed to outperform fixed-alphabet two-way QSDC on standard rate metrics. On a bits-per-pair basis, the protocol is upper-bounded by $\log_2 M / N$, which is strictly less than the $2$ bits per pair achievable in principle by ideal Bell-state encoding for any $N \geq 2$ and any $M$ consistent with the separability rule of Eq.~\eqref{eq:deltaP_spacing}. On a bits-per-second basis, the same gap persists at any fixed pair-detection rate. The protocol does, however, support a larger number of bits per symbol than fixed four-state Bell encoding for any $M > 4$, and it introduces protocol-level design parameters (alphabet size, integration window, separability margin) that are not available in fixed-alphabet schemes. The contribution of the present work should therefore be understood as a different operating point in the design space of two-way QSDC, rather than as a rate-competitive replacement for existing schemes. The intended use cases are operating environments in which the design flexibility, hardware simplicity, or analog-modulation capability offered by this approach outweigh the rate cost.

\subsection{Comparison to fixed four-state Bell-state encoding}
Within the broader class of two-way QSDC protocols, fixed four-state Bell-state encoding is the most direct comparison point. The present protocol differs from that family in four practical respects. First, the alphabet size $M$ is a protocol-level design parameter rather than a fixed property of the encoding, so the same hardware can be operated at different alphabet sizes to match application requirements or channel conditions. Second, the receiver does not perform a full Bell-state measurement and instead uses standard polarization-resolved coincidence detection, which avoids the engineering complexity of deterministic Bell-state measurement. Third, the protocol degrades gracefully under loss because missing pairs reduce the effective sample size of the histogram rather than producing symbol erasures, which distributes the impact of loss across many pairs rather than concentrating it into individual symbol erasures. Fourth, the encoded parameter $\theta$ can be modulated continuously in time, which provides a native pathway for transmitting analog waveforms without prior digitization. As previously indicated, none of these properties constitute a rate advantage over fixed-alphabet Bell-state encoding, and the present protocol is not proposed as a universal replacement for that approach. Instead, these properties make the protocol attractive in operating regimes where the corresponding operational benefits matter more than maximum per-pair rate.

\subsection{Relation to additional prior-art families}
Beyond fixed four-state Bell-state encoding discussed above, two additional families of prior work bear comparison to the protocol proposed here.

\emph{Continuous-variable QSDC.} CV-QSDC schemes encode information in the quadratures of bosonic modes, in the form of coherent, squeezed, or Gaussian-modulated states, read out by homodyne or heterodyne detection~\cite{Srikara2020CVQSDC,Cao2023CVQSDC,paparelle2025experimental,Gaussian2025CVQSDC}. The encoded parameter in those schemes is continuous and the carrier is the quantum state itself. The present scheme also employs a continuous encoding parameter, but the carrier is a polarization-entangled qubit pair in a finite-dimensional Hilbert space, and the continuous parameter modulates a classical probability distribution over discrete joint outcomes rather than a quantum quadrature. The decoding front-end is single-photon coincidence counting, not homodyne detection. The CV-QSDC family and the present scheme share the high-level idea of analog modulation but differ in the underlying quantum system, the modulation mechanism, and the receiver architecture.

\emph{High-dimensional / qudit QSDC.} A separate body of work enlarges the per-pair information capacity by enlarging the Hilbert-space dimension of the carrier through degrees of freedom such as time-bin, orbital angular momentum, or additional entangled degrees of freedom (hyperentanglement)~\cite{Wang2005HighDim,Cozzolino2019HighDim,HighDimSciRep2024,zeng2024high}. The alphabet is correspondingly enlarged to up to $d^2$ high-dimensional unitaries (analogous to the four Bell states for $d=2$). The present protocol does not enlarge the information-bearing Hilbert space. Information is encoded entirely within the two-qubit polarization degree of freedom. The polarization-entangled pair source used in this work is, in practical implementations, hyperentangled in polarization and one auxiliary degree of freedom (frequency or path) to enable deterministic separation of the two photons, but the auxiliary degree of freedom carries no encoded information and is used solely as a sorting label. The present protocol enlarges the effective alphabet by increasing the number of pairs $N$ accumulated per symbol, since the maximum reliably distinguishable alphabet size grows with $N$ as governed by the separability rule of Eq.~\eqref{eq:deltaP_spacing}. This temporal accumulation is a classical-statistics use of time, distinct from time-bin encoding in which time is used as an additional quantum degree of freedom of each photon. The two strategies are orthogonal: high-dimensional QSDC obtains capacity through hardware complexity in the source and receiver for the information-bearing degree of freedom, while the present scheme obtains capacity through temporal integration on simple polarization-resolved hardware and a fixed two-qubit information space.

\subsection{Implementation considerations and known two-way attack surfaces}
Two-way direct-communication protocols admit active attack vectors beyond passive interception, including man-in-the-middle/impersonation strategies and injected-light/Trojan-horse probing of Bob's apparatus~\cite{Wojcik2003PingPongEavesdrop,cai2003ping,Beaudry2013TwoWaySecurity,Pavicic2021MITMTwoWay,Gisin2006TrojanHorse}. This work does not claim new countermeasures to these known classes. Practical deployments should adopt standard mitigation strategies from the literature, including strong authentication of classical coordination, randomized test rounds/control modes, optical isolation and filtering, and monitoring detectors to detect injected signals~\cite{Gisin2006TrojanHorse,Pan2024EvolutionQSDC}.

\subsection{Scope and future work}
The present manuscript develops the encoding/decoding mechanism and provides a quantitative framework for bounding leakage under non-ideal effects. A full implementation-level security analysis can be developed by combining (i) test-mode disturbance statistics for active attacks with (ii) experimentally bounded leakage metrics for the traveling subsystem as described in Sec.~\ref{sec:leakage}, and by incorporating realistic loss, detector models, and finite-size statistics. In a related direction, the leakage analysis presented here is anchored to pairwise trace-distance diagnostics evaluated on a representative parametric channel model. A more general theoretical treatment, in which closed-form expressions for the $N_E$-copy trace distance, the full $M$-ary Holevo quantity, and broader families of non-ideal channel maps are developed without reliance on a specific parametric form, is left as an opportunity for future work. Such a treatment would establish tight rather than illustrative bounds and would extend the framework to cover non-diagonal reduced states and correlated noise models that fall outside the scope of the present analysis. A distinct future direction is to explore non-histogram decoding strategies for the same encoded states $\ket{\psi(\theta)}$. The present protocol reconstructs joint-outcome histograms over a window of $N$ pairs and identifies the closest symbol by comparison with template histograms. Alternative approaches include adaptive measurement strategies, in which the measurement basis on later pairs is chosen based on the outcomes of earlier pairs in the same window, and maximum-likelihood decoding, which computes the likelihood of each candidate $\theta_m$ directly from the recorded outcomes rather than from a reconstructed histogram. Such strategies cannot exceed the Holevo-bounded ceiling of $2$ bits per pair set by the two-qubit Hilbert space, but may achieve better symbol distinguishability at smaller $N$ than the present histogram-based approach requires. The development of these alternative decoding strategies, and the security analysis appropriate to each, is left to future work.

\bibliographystyle{apsrev4-2}
\bibliography{references}

\end{document}